%Paper: hep-ph/9407207
%From: Kazunori_Takenaga <take1scp@mbox.nc.kyushu-u.ac.jp>
%Date: Sat, 2 Jul 94 17:11:05 JST

%% -----------------------------------------------------------------------
% Two PostScript figures ( fig1.ps and fig2.ps ) are appended at the end
% of the text ( uuencoded tar.Z ). Send fig1.ps and fig2.ps separately to
% a PostScript printer.
%
% Please choose l ( reduced mode ) in harvmac.
%% -----------------------------------------------------------------------
\input harvmac

\def\NP#1{Nucl.\ Phys. {\bf B{#1}}}
\def\PL#1{Phys.\ Lett. {\bf B{#1}}}
\def\PR#1{Phys.\ Rev. {\bf {#1}}}
\def\PRD#1{Phys.\ Rev. {\bf D{#1}}}

\def\PRL#1{Phys.\ Rev.\ Lett. {\bf {#1}}}
\def\PTP#1{Prog.\ Theor.\ Phys. {\bf {#1}}}

\overfullrule=0pt
\catcode `\@=11
\newfam\mibfam

\font\tenmib=cmmib10
\font\sevenmib=cmmib7
\font\fivemib=cmmib5
\skewchar\tenmib='177
\skewchar\sevenmib='177
\skewchar\fivemib='177
\textfont\mibfam=\tenmib
\scriptfont\mibfam=\sevenmib
\scriptscriptfont\mibfam=\fivemib
\let\rel@x=\relax
\let\n@expand=\relax
\newcount\f@ntkey  \f@ntkey=0
\def\f@m{\afterassignment\samef@nt\f@ntkey=}
\def\samef@nt{\fam=\f@ntkey \the\textfont\f@ntkey\rel@x}
\def\rm{\n@expand\f@m0 }
\def\mit{\n@expand\f@m1 }
\def\cal{\n@expand\f@m2 }
\def\it{\n@expand\f@m\itfam}
\def\sl{\n@expand\f@m\slfam}
\def\bf{\n@expand\f@m\bffam}
\def\tt{\n@expand\f@m\ttfam}
\def\mib{\n@expand\f@m\mibfam}
\def\lsim{\ \raise.3ex\hbox{$<$\kern-.75em\lower1ex\hbox{$\sim$}}\ }
\def\gsim{\ \raise.3ex\hbox{$>$\kern-.75em\lower1ex\hbox{$\sim$}}\ }
\def\gl{\ \raise.5ex\hbox{$>$}\kern-.8em\lower.5ex\hbox{$<$}\ }

\catcode `\@=12
%%%%%%%%%%%%%%%%%%%%%%%%%%%%%%%%%%%%%%%%
\def\e{{\rm e}}
\def\del{\partial}
\def\dslash{\del\kern-0.55em\raise 0.14ex\hbox{/}}

\def\bfx{{\mib x}}
\def\bfpT{{{\mib p}_T}}
\def\abs#1{{\left|{#1}\right|}}
\def\expecv#1{\langle #1 \rangle}
\def\phipa{\phi^{(+\alpha)}}
\def\phima{\phi^{(-\alpha)}}

\def\a{\alpha}
\def\b{\beta}
\def\ep{\epsilon}
\def\g{\gamma}

\def\s{\sigma}
\def\t{\tau}
\def\Fig{Fig.~\the\figno\nfig}
%%%%%%%%%%%%%%%%%%%%%%%% Title page %%%%%%%%%%%%%%%%%%%%%%%%%%%%%%%%%%%%%%%%%

% Title
\Title{\vbox{\baselineskip12pt\hbox{KYUSHU-HET-18}\hbox{SAGA-HE-64}\hbox{July
1994}}}
{\vbox{\centerline{Fermion Scattering off a CP-Violating Electroweak Bubble
Wall II}}}
\centerline{Koichi
Funakubo${}^{1)}$\footnote{$^a$}{funakubo@sagagw.cc.saga-u.ac.jp}, Akira
Kakuto${}^{2)}$\footnote{$^b$}{ kakuto@fuk.kindai.ac.jp}, Shoichiro
Otsuki${}^{2)}$}
\centerline{Kazunori Takenaga${}^{3)}$
\footnote{$^c$}{take1scp@mbox.nc.kyushu-u.ac.jp}  and
Fumihiko Toyoda${}^{2)}$\footnote {$^d$}{ftoyoda@fuk.kindai.ac.jp}}
\vskip 0.5cm

\centerline{${}^{1)}${\it Department of Physics, Saga University, Saga, 840
JAPAN}}
\smallskip
\centerline{${}^{2)}${\it Department of Liberal Arts, Kinki University in
Kyushu, Iizuka 820 JAPAN}}
\smallskip
\centerline{${}^{3)}${\it Department of Physics, Kyushu University, Fukuoka,
812 JAPAN}}

% Abstract
\vskip 0.5cm
By treating CP-violating interaction as a perturbative term,
we solved in a previous paper the Dirac equation in the background
of electroweak bubble wall, and obtained
the transmission and
reflection coefficients for a chiral fermion incident from the
symmetric-phase region.
We give the transmission and reflection coefficients
under the other boundary condition, that is, the case of the fermion
incident from the broken-phase region.
There hold the respective sets of unitarity relations and
also reciprocity relations among them. These relations enable us to
obtain a simple form of  quantum-number flux through the bubble wall, which
is the first order quantity of the CP violation.
A factor in the integrand of the flux, which originates from the CP violation,
is
very sensitive to the functional form of the CP-violating phase. The absolute
value of
this factor is found to decrease as $m_0/a$ increases for $m_0/a
{}~\roughly{>}~1$,
where $1/a$ is the wall thickness and $m_0$ the fermion mass near the
critical temperature.

\Date{6/94}

%\draft % for prerimanary versions or specify \draftmode at some point.

%%%%%%%%%%%%%%%%%%%%%% The title page ends. %%%%%%%%%%%%%%%%%%%%%%%%%%%%%%%%%
%%%%%%%%%%%%%%%%%%%%%%%%%%%%%%%%%%%%%%%%%%%%%%%%%%%%%%%%%%%%%%%%%%%%%%%%%%%%%%%

%%%%%%%%%%%%%%%%%%%%%% Body %%%%%%%%%%%%%%%%%%%%%%%%%%%%%%%%%%%%%%%%%%%%%%%%%
%\baselineskip=0.5 truein plus 2pt minus 1pt
\newsec{Introduction}
Electroweak baryogenesis is one of the most challenging problems in
particle physics and cosmology. The standard model meets the three requirements
to generate baryon number \ref\sak{A. Zakharov, JETP Lett. {\bf 5} (1967) 24.}
, provided that the phase transition is of first order
\ref\krs{V. Kuzmin, V. Rubakov and M. Shaposhnikov, \PL{155} (1985) 36.}.
One of them, the baryon-number violating process is supplied by the chiral
anomaly\ref\tooft{G. 'tHooft, \PRL{37} (1976) 8.}. At high temperature, the
rate of
the anomalous baryon- and lepton-number violating
processes induced by the sphaleron may be rapid enough to be in thermal
equilibrium,
while the rate becomes negligible at low temperature.
The second requirement, out-of-thermal-equilibrium condition, could
be realized by the first-order phase transition.
The bubble nucleating in the symmetric-phase region expands to convert the
whole
of the universe
into the broken phase.
In the minimal standard model or its extension, the phase transition
can be of first order by choice of parameters in the Higgs sector, as discussed
by many
authors
\ref\dl{M. Dine, P. Huet, A. Linde and D. Linde, \PR{46} (1992) 550.}
\ref\mcl{B. Liu, L. McLerran and N. Turok, \PR{46} (1992) 2668.}
\ref\turok{N. Turok and J. Zadrozny, \NP{369} (1992) 729.}
\ref\fkt{K. Funakubo, A. Kakuto and K. Takenaga, \PTP{91} (1994) 341.}.
The last requirement of CP violation is given by
the Kobayashi-Maskawa scheme
\ref\km{M. Kobayashi and T. Maskawa, \PTP{49} (1973) 652.} in the minimal
standard
model. The effect of CP violation in this scheme, however,  may arise
in higher orders
\ref\fs{G. R. Farrar and M. E. Shaposhnikov, \PRL{70} (1993) 2833;
                   Preprint CERN-TH-6734/1993---RU-93-11.}
\ref\orloff{M. B. Gavela, P. Hernandez, J. Orloff and O. P\`ene, Mod. Phys.
Lett.{\bf 9}
(1994) 795.}
, so that its influence on
baryogenesis would inevitably be too small to explain the baryon asymmetry
observed today
\ref\huet{P. Huet and E. Sather, SLAC-PUB-6479.}.
Then, one may need a new source of CP violation which would
be realized in multi-Higgs models.
Thus, a moving bubble wall with CP violation formed by development of the VEV
of
the Higgs fields would be crucial for electroweak baryogenesis. \par
In a previous paper\ref\foktt{K. Funakubo, A. Kakuto, S. Otsuki, K. Takenaga
and F.Toyoda,
KYUSHU-HET-15---SAGA-HE-55, to be puplished in \PRD{50} (1994).}
(referred to as I hereafter)
we presented a general prescription to study the fermion propagation
in CP-violating bubble-wall background.
By regarding the CP-violating term as a
perturbation, we solved the full equation to the first order of the CP
violation by DWBA ( the distorted-wave Born approximation ) method.
We derived transmission and reflection
coefficients of fermions from the asymptotic form of the wave function
with the boundary condition that an incident fermion comes from the symmetric
phase.
\par
In this paper we study the other case where an incident fermion comes from the
broken phase.
We derive the transmission and reflection coefficients and
establish relations between quantities derived in this paper and those in I.
We show that reciprocity relations between the two cases hold, at least, up
to the first order of CP violation.
The reciprocity relations enable us to simplify quantum-number
flux through the bubble wall. We find that the quantum-number flux is
proportional
to $(Q_L -Q_R)\Delta R$.
Here $Q_L(Q_R)$ is a quantum number carried by the
left ( right )-handed fermion, $\Delta R \equiv R_{R \rightarrow L}^s
- \bar R_{R \rightarrow L}^s$ and
$R_{R \rightarrow L}^s (\bar R_{R \rightarrow L}^s)$ is
the reflection coefficient for right-handed chiral fermion ( antifermion )
incident from the symmetric phase.
We analyze $\Delta R$ numerically \ref\fotr{Because of a trivial error
in our program for numerical calculation, figures and table in \foktt~ are
incorrect.}
by taking various forms of CP-violating profile of the bubble wall.
The quantity $\Delta R$ is very sensitive to the
functional form of the CP-violating phase. Our numerical analyses show that
$\abs{\Delta R}$ decreases as $m_0/a$ increases when
$1/m_0 ~\roughly{<}~1/a$, where $1/a$ is the thickness of the bubble wall and
$m_0$
is the fermion mass \ref\fotno{Here the fermion mass $m_0$ is determined by
the VEV of the Higgs field
near the critical temperature. See Eq. (2.2).}.
This means that the heavier fermion gives the smaller contribution to
$\abs{\Delta R}$
for fixed $a$, if the Compton wave length is comparable with or less than the
wall thickness.
One should note that~{\it a larger Yukawa coupling does not necessarily mean a
larger
effect of CP violation in electroweak baryogenesis}. \par
%
%%%%%%%%%%%%%%%%%%%%%%%%%%%%%%%%%%%%%%
This paper is organized as follows. We give a summary of I in section 2.
In section 3, we present the transmission and reflection
coefficients for a fermion incident from the broken phase. We also prove the
reciprocity
relations to the first order of the CP violation.
In section 4 we obtain a simple form of quantum-number flux into the symmetric
phase.
In the final section we discuss the physical
CP-violating phase in the Higgs sector and give concluding remarks.
%% FOLLOWING LINE CANNOT BE BROKEN BEFORE 80 CHAR
%%%%%%%%%%%%%%%%%%%%%%%%%%%%%%%%%%%%%%%%%%%%%%%%%%%%%%%%%%%%%%%%%%%%%%%%%%%%%%%%%%%%%%%
\newsec{DWBA to CP-Violating Dirac Equation---The Case of Fermion
Incident from Symmetric-Phase Region---}
In this section we
briefly describe how to treat fermion propagation in the bubble-wall background
by regarding CP-violating term as a perturbation. The transmission and
reflection
coefficients for a fermion incident from the symmetric phase are also given.
\subsec{Dirac equation and ansatz}
We consider one-flavor model described by the lagrangian,
\eqn\aa{
  \CL = \bar\psi_L i\dsl\psi_L + \bar\psi_R i\dsl\psi_R
         + ( f \bar\psi_L\psi_R\phi + {\rm h.c.} ).
}
In the vacuum, near the first-order phase transition,
$\expecv\phi$ may be $x$-dependent field, so that we put
\eqn\ab{
  m(\bfx) = - f \expecv\phi(\bfx),
}
where $m(\bfx)$ is complex-valued and we neglect the time dependence. If the
phase of $m(\bfx)$ has no spatial dependence, it is removed by a constant
bi-unitary transformation, which is outside of our interest.
The Dirac equation describing fermion propagation in the bubble-wall
background  with CP violation is
\eqn\ac{
i\dsl \psi(t,\bfx)-m(\bfx)P_R\psi(t,\bfx)-m^*(\bfx)P_L\psi(t,\bfx) = 0.
}
For the bubble wall with large enough radius, $m(\bfx)$ could be regarded as
a function of only one spatial coordinate, so that we put
$m(\bfx)=m(z)$.\par
To solve \ac~, we take the following ansatz:
\eqn\ad{\eqalign{
            \psi(t,\bfx)
 &= (i\dsl  + m^*(z)P_R + m(z)P_L)
    \e^{i\s(-Et+\bfpT\bfx_T)}\psi_E(\bfpT,z) \cr
 &= \e^{i\s(-Et+\bfpT\bfx_T)}  \cr
 &\times [\s(\g^0E-\g_Tp_T)
    + i\g^3\del_z+ m^*(z)P_R + m(z)P_L] \psi_E (\bfpT,z), \cr
}
}
where $\s=+(-)$ for positive (negative)-energy solution,
$\bfpT=(p^1,p^2)$, $\bfx_T=(x^1,x^2)$, $p_T=\abs{\bfpT}$ and
$\g_Tp_T=\g^1p^1+\g^2p^2$. By putting $E=E^*\cosh\eta$ and
$p_T=E^*\sinh\eta$ with $E^*=\sqrt{E^2-p_T^2}$, the Lorentz
transformation eliminates $\bfpT$.
After this Lorentz rotation for a fixed $\bfpT$,
the Dirac equation is rewritten as
\eqn\ae{
 \bigl[ {E^*}^2 + \del_z^2-{\abs{m(z)}}^2
    +im_R^\prime(z)\g^3 - m_I^\prime(z)\g_5\g^3 \bigr]\psi_E(z)=0,
}
where $m(z)=m_R(z)+im_I(z)$. Now let us introduce a set of dimensionless
variables using a parameter $a$, whose inverse characterizes the thickness
of the wall: $m_R(z)=m_0f(az)=m_0f(x)$, $m_I(z)=m_0g(az)=m_0g(x)$,
$x\equiv az$, $\t\equiv at$, $\ep \equiv E^*/a$ and $\xi\equiv m_0/a$,
where $m_0$ is the fermion mass in the broken phase. Eq.\ae~ is expressed as
\eqn\af{
 [\ep^2 + {d^2\over{dx^2}} - \xi^2(f(x)^2+g(x)^2)
  +i\xi f^\prime(x)\g^3-\xi g^\prime(x)\g_5\g^3]
  \psi_\ep(x) = 0.
}
\par
As for $f(x)$ and $g(x)$, we do not specify their functional
forms but only assume that
\eqn\ag{
 f(x)\rightarrow\cases{1, &as $x\rightarrow+\infty$,\cr
                       0, &as $x\rightarrow-\infty$,\cr}
}
and that $\abs{g(x)}<<1$, ${\it {i.e.,}}$ small CP violation.
Eq.\ag~  means that the system is in the broken (symmetric) phase at
$x\sim+\infty$ ($x\sim-\infty$), the wall height being $m_0$.
\par

\subsec{DWBA to the Dirac equation}
We regard the small $\abs{g(x)}$ as a perturbation, and keep quantities
up to $O(g^1)$. Put
\eqn\ah{
   \psi_\ep(x) = \psi^{(0)}(x) + \psi^{(1)}(x),
}
where $\psi^{(0)}(x)$ is a solution to the unperturbed equation
\eqn\ai{
 [\ep^2 + {d^2\over{dx^2}} - \xi^2f(x)^2+i\xi f^\prime(x)\g^3]
  \psi^{(0)}(x) = 0
}
with an appropriate boundary condition. Then $\psi^{(1)}(x)$ of $O(g^1)$
is solved as
\eqn\aj{
 \psi^{(1)}(x) = \int dx^\prime G(x,x^\prime)V(x^\prime)
                 \psi^{(0)}(x^\prime)\qquad {\rm with}\qquad
   V(x) = -\xi g^\prime(x)\g_5\g^3.
}
$G(x,x^\prime)$ is the Green's function for the operator in \ai~ satisfying
the same boundary condition as $\psi^{(0)}(x)$. To this order, the solution
to the Dirac equation is given by
\eqn\ak{
 \psi(x)_{\sigma}\simeq \e^{-i\s\ep\t}\Bigl\{ [\s\ep\g^0 +
    i\g^3{d\over{dx}}+\xi f(x)][\psi^{(0)}(x) + \psi^{(1)}(x)]
         - i\xi g(x)\g_5\psi^{(0)}(x) \Bigr\}.
}
If we expand $\psi^{(0)}(x)$ in terms of the eigenspinors of $\g^3$
as $\psi^{(0)}(x)\sim \phi_\pm(x)u^s_\pm$ with
$ \gamma^3 u^s_\pm = \pm i u^s_\pm (s=1,2)$, $\phi_\pm(x)$ satisfies
\eqn\al{
  [\ep^2 + {d^2\over{dx^2}} - \xi^2f(x)^2 \mp \xi f^\prime(x)]
  \phi_\pm(x) = 0.
}
Because of \ag~, the asymptotic forms of $\phi_\pm(x)$ should be
$ \phi_\pm(x)\rightarrow \e^{\a x}, \e^{-\a x} (x\rightarrow+\infty)$ and
$\e^{\b x}, \e^{-\b x} (x\rightarrow-\infty),$
where $\a=i\sqrt{\ep^2-\xi^2}$ and $\b=i\ep$. Putting all these together,
we can obtain the asymptotic forms of the wave function \ak~ at
$x \rightarrow \pm \infty$.
\subsec{Fermion incident from symmetric-phase region}
We consider a state in which the incident wave coming from $x=-\infty$ is
reflected in part at the bubble wall, while at $x=+\infty$ only the
transmitted wave exists. We denote two independent solutions to \al~
as $\phipa_\pm(x)$ and $\phima_\pm(x)=\bigl(\phipa_\pm(x)\bigr)^*$ which
behave as
\eqn\am{\eqalign{
 \phipa_\pm(x)\rightarrow\e^{\a x},\qquad
 \phima_\pm(x)\rightarrow\e^{-\a x}
}
}
at $x\rightarrow+\infty$.
Their asymptotic forms at $x\rightarrow-\infty$ are
\eqn\an{\eqalign{
 &\phipa_\pm(x)\sim \g_\pm(\a,\b)\e^{\b x}+\g_\pm(\a,-\b)\e^{-\b x}, \cr
 &\phima_\pm(x)\sim \g_\pm(-\a,\b)\e^{\b x}+\g_\pm(-\a,-\b)\e^{-\b x}.\cr
}
}
{}From these, the general solution to \ai~ is eventually given as \foktt
\eqn\ao{
 \psi^{(0)}(x) = \sum_s
 [A_s^{(-)}\phima_+(x) + A_s^{(+)}\phipa_+(x)]u^s_+.
}
The required boundary condition is achieved by setting $A_s^{(-)}=0$
for $\s=+$ and $A_s^{(+)}=0$ for $\s=-$.
The Green's function which matches this boundary condition can be
constructed from $\phi^{(\pm\a)}_\pm(x)$\foktt.
\par
{}From the asymptotic forms of
$\bigl(\psi(x)_{\s}\bigr)^{inc}$,
$\bigl(\psi(x)_{\s}\bigr)^{trans}$ and
$\bigl(\psi(x)_{\s}\bigr)^{refl}$ for $\s=\pm$ in \ak~\foktt\  ,
we obtain those of the vector and axial-vector currents,
$j_V^\mu = \bar\psi\g^\mu\psi$ and $j_A^\mu = \bar\psi\g^\mu\g_5\psi$.
In terms of the chiral currents, $j^\mu_L = (1/2)(j^\mu_V-j^\mu_A)$ and
$j^\mu_R = (1/2)(j^\mu_V+j^\mu_A)$,
the transmission and reflection coefficients for the chiral fermion
are defined as
\eqn\ap{\eqalign{
 &T^{(\s)}_{L\rightarrow L(R)}
 = \bigl(j^3_{L(R),\s}\bigr)^{trans}\bigr|_{A_1^\s=0}
  /\bigl(j^3_{L,\s}\bigr)^{inc},        \cr
 &T^{(\s)}_{R\rightarrow L(R)}
 = \bigl(j^3_{L(R),\s}\bigr)^{trans}\bigr|_{A_2^\s=0}
  /\bigl(j^3_{R,\s}\bigr)^{inc},         \cr
 &R^{(\s)}_{L(R)\rightarrow R(L)}
 = -\bigl(j^3_{R(L),\s}\bigr)^{refl}
   /\bigl(j^3_{L(R),\s}\bigr)^{inc}.        \cr
}
}
If we denote $R^s(T^s)=R^{(+)}(T^{(+)})$ and
$\bar R^s(\bar T^s)=R^{(-)}(T^{(-)})$,
where the superscript $s$ denotes the fermion incident from the
symmetric phase, we have\ref\fotd{In this paper we denote $\delta^{inc}$ in I
as
$\delta^{CP}$. }
\eqn\aq{\eqalign{
 &T_{L\rightarrow L}^s=\bar T_{R\rightarrow R}^s
=
T^{(0)} \bigl({1\over 2}+{\ep \over {2\sqrt{\ep^2-\xi^2}}}\bigr)
 (1-\delta^{CP}),     \cr
 &T_{L\rightarrow R}^s=\bar T_{R\rightarrow L}^s
=
T^{(0)}\bigl({1\over 2}-{\ep \over {2\sqrt{\ep^2-\xi^2}}}\bigr)
 (1-\delta^{CP}),     \cr
 &T_{R\rightarrow L}^s=\bar T_{L\rightarrow R}^s
=
T^{(0)}\bigl({1\over 2}-{\ep \over {2\sqrt{\ep^2-\xi^2}}}\bigr)
 (1+\delta^{CP}),     \cr
 &T_{R\rightarrow R}^s=\bar T_{L\rightarrow L}^s
=
T^{(0)}\bigl({1\over 2}+{\ep \over {2\sqrt{\ep^2-\xi^2}}}\bigr)
 (1+\delta^{CP}),     \cr
 &R_{L\rightarrow R}^s=\bar R_{R\rightarrow L}^s
=R^{(0)}+T^{(0)}\delta^{CP}
 ,   \cr
 &R_{R\rightarrow L}^s=\bar R_{L\rightarrow R}^s
=R^{(0)}-T^{(0)}\delta^{CP}.    \cr
}
}
Here
\eqn\ar{\eqalign{
T^{(0)}
={\alpha \over \beta}{1 \over {\abs{\gamma_{+}(\alpha, \beta)}}^2},\qquad
R^{(0)}
=\abs{{{\gamma_{+}(\alpha, -\beta)} \over {\gamma_{+}(\alpha,
\beta)}}}^2
}
}
with $T^{(0)}+R^{(0)}=1$ are respectively the transmission and reflection
coefficients in the absence of CP violation.
The correction by CP violation is
\eqn\as{\eqalign{
 \delta^{CP} = {\xi\over{2\sqrt{\ep^2-\xi^2}}}
 \Bigl( {\g_-(-\a,\b)\over\g_+(\a,\b)}I_2 + c.c. \Bigr),
}
}
where
\eqn\at{ I_2\equiv\int_{-\infty}^{\infty}dx\,
g^\prime(x)\phipa_-(x)\phipa_+(x).
}
Among these, the following unitarity relations hold:
\eqn\au{\eqalign{
 T^s_{L\rightarrow L}+T^s_{L\rightarrow R}+R^s_{L\rightarrow R}=1, \qquad
 T^s_{R\rightarrow L}+T^s_{R\rightarrow R}+R^s_{R\rightarrow L}=1.
}
}

\newsec{Fermion Incident from Broken-Phase Region and Reciprocity}
\subsec{Asymptotic forms of the wave function}
In place of $\phipa_\pm(x)$, we start with $\phi^{(-\beta)}_\pm(x)$ that behave
as
\eqn\ba{\eqalign{
\phi^{(-\beta)}_\pm(x)\rightarrow\e^{-\beta x},\qquad
\phi^{(+\beta)}_\pm(x)=\bigl(\phi^{(-\beta)}_\pm(x)\bigr)^*
\rightarrow\e^{+\beta x}
}
}
at $x\rightarrow-\infty$, while
at $x\rightarrow+\infty$
\eqn\bb{\eqalign{
 &\phi^{(-\b)}_\pm(x)\sim {\widetilde \g}_\pm(-\b,\a)\e^{\a x}+{\widetilde
\g}_\pm(-\b,-\a)\e^{-\a x}, \cr
 &\phi^{(+\b)}_\pm(x)\sim {\widetilde \g}_\pm(\b,\a)\e^{\a x}+{\widetilde
\g}_\pm(\b,-\a)\e^{-\a x}.\cr
}}
{}From these, the general solution to \ai~ is given as
\eqn\bc{
 \psi^{(0)}(x) = \sum_s
 [B_s^{(-)}\phi_+^{(-\b)}(x) + B_s^{(+)}\phi_+^{(+\b)}(x)]u^s_+.
}
The boundary condition is achieved by setting $B_s^{(+)}=0$
for $\s=+$ and $B_s^{(-)}=0$ for $\s=-$.
\par
Most of the following formulae are derived completely in parallel to I,
so we skip the detailed algebra.
The asymptotic forms for the positive-energy wave are
\eqn\bd{\eqalign{
\bigl(\psi(x)_{\sigma=+}\bigr)^{trans}&={\rm e}^{-i \epsilon \tau -
\beta x}
                   \sum_s B_s ^{(+)}\bigl\{ \beta \big[1-(-)^s {\xi \over {2
\epsilon}}
                          ({\tilde I_1} - g(+\infty)) \bigr]u_{+}^s  \cr
                    & +\epsilon \bigl[1-(-)^s {\xi \over \epsilon}\bigl({1
\over 2}{\tilde I_1} -
                      {1 \over 2}g(+\infty) +g(-\infty) \bigr) \bigr] u_{-}^s
                                       \bigr\},\cr
\bigl(\psi(x)_{\sigma=+}\bigr)^{inc}&={\rm e}^{-i \epsilon \tau - \alpha
x}
                          {\tilde \gamma_{+}(-\beta, -\alpha)}\sum_s B_s^{(+)}
\cr
                    &\times \big\{ (\xi + \alpha) \big[1+(-)^s {\xi \over
\epsilon}
                     \bigl({{\xi -\alpha} \over {2 \beta}}
{{\tilde \gamma_{-}(\beta, -\alpha)} \over {\tilde \gamma_{+}(-\beta,
-\alpha)}}
{\tilde I_2} + {1 \over 2}g(+\infty) \bigr) \big] u_{+}^s \cr
                  &+ \epsilon[1 + (-)^s {\xi \over \epsilon}
                \bigl({{\xi - \alpha} \over {2\beta}}
{{\tilde \gamma_{-}(\beta, -\alpha)} \over {\tilde \gamma_{+}(-\beta,
-\alpha)}}
            {\tilde I_2} -{1 \over 2}g(+\infty) \bigr) \big] u_{-}^s \big\},
\cr
\bigl(\psi(x)_{\sigma=+}\bigr)^{refl}&={\rm e}^{-i \epsilon \tau +
\alpha x}
{\tilde \gamma_{+}(-\beta, \alpha)}\sum_s B_s^{(+)}  \cr
                  &\times \big\{ (\xi - \alpha) \big[1+(-)^s {\xi \over
\epsilon}
                 \bigl({{\xi +\alpha} \over {2 \beta}}
          {{\tilde \gamma_{-}(\beta,\alpha)} \over {\tilde \gamma_{+}(-\beta,
\alpha)}}
           {\tilde I_2} + {1 \over 2}g(+\infty) \bigr) \big] u_{+}^s \cr
              &+ \epsilon[1 + (-)^s {\xi \over \epsilon}
            \bigl({{\xi + \alpha} \over {2\beta}}
          {{\tilde \gamma_{-}(\beta, \alpha)} \over {\tilde \gamma_{+}(-\beta,
\alpha)}}
              {\tilde I_2} -{1 \over 2}g(+\infty) \bigr)\big] u_{-}^s \big\},
\cr
}
}
and
\eqn\be{
\big(\psi(x)_{\sigma=-}\big)^{asym}=
\big(\psi(x)_{\sigma=+}\big)^{asym}|_{(\alpha,
\beta,\epsilon, {\tilde I_1}, {\tilde I_2})\rightarrow
(-\alpha, -\beta, -\epsilon, {\tilde I_1}^*, {\tilde I_2}^*)}
}
for negative-energy wave. ${\tilde I_1}$ and ${\tilde I_2}$ are defined by
\eqn\be{\eqalign{
{\tilde I_1}&\equiv \int_{-\infty}^{+\infty}dx
g^{\prime}(x)\phi_{-}^{(+\beta)}(x)\phi_{+}^{(-\beta)}(x),\cr
{\tilde I_2}&\equiv \int_{-\infty}^{+\infty}dx
g^{\prime}(x)\phi_{-}^{(-\beta)}(x)\phi_{+}^{(-\beta)}(x).\cr
}
}
The first one ${\tilde I_1}$ is found to be irrelevant to the final results
similarly to
$I_1\equiv \int_{-\infty}^{+\infty}dx
g^{\prime}(x)\phi_{-}^{(-\alpha)}(x)\phi_{+}^{(+\alpha)}(x)$ in I.
\subsec{Transmission and reflection coefficients}
These are defined in the way similar to \ap~, and are
expressed as follows:
\eqn\bg{\eqalign{
&T_{L\rightarrow L}^b ={\bar T_{R\rightarrow R}}^b
={\tilde T^{(0)}} \bigl({1 \over 2}+{{\sqrt{\epsilon^2 - \xi^2}} \over {2
\epsilon}}
\bigr)
(1 + \tilde \delta^{CP}),      \cr
&T_{L\rightarrow R}^b ={\bar T_{R\rightarrow L}}^b
={\tilde T^{(0)}} \bigl({1 \over 2}-{{\sqrt{\epsilon^2 - \xi^2}} \over {2
\epsilon}}
\bigr)
(1 - \tilde \delta^{CP}),      \cr
&T_{R\rightarrow L}^b ={\bar T_{L\rightarrow R}}^b
={\tilde T^{(0)}} \bigl({1 \over 2}-{{\sqrt{\epsilon^2 - \xi^2}} \over {2
\epsilon}}
\bigr)
(1 + \tilde \delta^{CP}),      \cr
&T_{R\rightarrow R}^b ={\bar T_{L\rightarrow L}}^b
={\tilde T^{(0)}} \bigl({1 \over 2}+{{\sqrt{\epsilon^2 - \xi^2}} \over {2
\epsilon}}
\bigr)
(1 - \tilde \delta^{CP}),      \cr
}
}
and
\eqn\bh{\eqalign{
&R_{L \rightarrow L}^b=\bar R_{R \rightarrow R}^b
=-R_{R \rightarrow R}^b=-\bar R_{L \rightarrow L}^b
={\tilde T^{(0)}}{\xi^2 \over {2 \epsilon \sqrt{\epsilon^2 - \xi^2}}}
\tilde \delta^{CP}, \cr
&R_{L \rightarrow R}^b = \bar R_{R \rightarrow L}^b ={\tilde R^{(0)}} -
{\tilde T^{(0)}}\bigl(
{\sqrt{\epsilon^2 - \xi^2} \over \epsilon} +
{\xi^2 \over {2 \epsilon \sqrt{\epsilon^2 - \xi^2}}} \bigr)
\tilde \delta^{CP},    \cr
&R_{R \rightarrow L}^b = \bar R_{L \rightarrow R}^b ={\tilde R^{(0)}} +
{\tilde T^{(0)}}\bigl(
{\sqrt{\epsilon^2 - \xi^2} \over \epsilon} +
{\xi^2 \over {2 \epsilon \sqrt{\epsilon^2 - \xi^2}}} \bigr)
 \tilde \delta^{CP},    \cr
}
}
where superscript $b$ denotes the fermion incident from the broken phase.
Note that, in contrast to the previous case, we have nonzero $R_{L(R)
\rightarrow L(R)}^b$ because
the fermion is massive in the broken phase.
${\tilde T^{(0)}}$ and ${\tilde R^{(0)}}$, the transmission and reflection
coefficients in the
absence of CP violation respectively, are given by
\eqn\bi{
{\tilde T^{(0)}}= {\beta \over \alpha}{1 \over {|\tilde \gamma_{+}(\beta,
\alpha)|^2}}, \quad
{\tilde R^{(0)}}=\abs{{\tilde \gamma_{+}(\beta, -\alpha)} \over {\tilde
\gamma_{+}(\beta, \alpha)}}^2,
}
with ${\tilde T^{(0)}}+{\tilde R^{(0)}}=1.$
The correction by CP violation is given by
\eqn\bj{\eqalign{
{\tilde \delta}^{CP}={\xi \over \epsilon}\bigl[{{\xi - \alpha} \over {2
\beta}}
{{\tilde \gamma_{-}(\beta , -\alpha)} \over {\tilde \gamma_{+}(-\beta,
-\alpha)}}{\tilde I_2}
+ c.c.\bigl].
}
}
Among these the following unitarity relations hold as they should:
\eqn\bk{\eqalign{
&T_{L \rightarrow L}^b + T_{L \rightarrow R}^b +
R_{L \rightarrow L}^b + R_{L \rightarrow R}^b =1,  \cr
&T_{R \rightarrow L}^b + T_{R \rightarrow R}^b +
R_{R \rightarrow L}^b + R_{R \rightarrow R}^b =1.  \cr
}
}
\subsec{Reciprocity relations}
$\phi_{\pm}^{(\pm\beta)}(x)$ are linearly dependent on
$\phi_{\pm}^{(\pm\alpha)}(x)$ and {\it vice versa}.
By comparing their asymptotic forms,
we have
\eqn\bl{\eqalign{
\phi_{\pm}^{(+\beta)}(x)&={\tilde
\gamma_{\pm}(\beta,\alpha)\phi_{\pm}^{(+\alpha)}(x)}+
{\tilde \gamma_{\pm}(\beta,-\alpha)\phi_{\pm}^{(-\alpha)}(x)},\cr
\phi_{\pm}^{(-\beta)}(x)&={\tilde
\gamma_{\pm}(-\beta,\alpha)\phi_{\pm}^{(+\alpha)}(x)}+
{\tilde \gamma_{\pm}(-\beta,-\alpha)\phi_{\pm}^{(-\alpha)}(x)}\cr
}}
and
\eqn\nh{\eqalign{
\phi_{\pm}^{(+\alpha)}(x)&={
\gamma_{\pm}(\alpha,\beta)\phi_{\pm}^{(\beta)}(x)}+
{ \gamma_{\pm}(-\alpha,\beta)\phi_{\pm}^{(-\beta)}(x)},\cr
\phi_{\pm}^{(-\alpha)}(x)&={\gamma_{\pm}(-\alpha,
\beta)\phi_{\pm}^{(+\beta)}(x)}+
{ \gamma_{\pm}(-\alpha,-\beta)\phi_{\pm}^{(-\beta)}(x)},\cr
}}
which yield relations between $\tilde \gamma$ and $\gamma$:
\eqn\bm{\eqalign{
&\tilde \gamma_{\pm}(\beta, \alpha)={\beta \over
\alpha}\gamma_{\pm}(-\alpha,-\beta), \quad
\tilde \gamma_{\pm}(\beta, -\alpha)={-\beta \over
\alpha}\gamma_{\pm}(\alpha,-\beta), \cr
&\tilde \gamma_{\pm}(-\beta, \alpha)={-\beta \over
\alpha}\gamma_{\pm}(-\alpha,\beta),
\tilde \gamma_{\pm}(-\beta, -\alpha)={\beta \over
\alpha}\gamma_{\pm}(\alpha,\beta). \cr
}}
Eq.\bm~ immediately leads us to
\eqn\bn{{\tilde T^{(0)}}=T^{(0)},\qquad {\tilde R^{(0)}}=R^{(0)},
}
that is, the reciprocity relations in the absence of CP violation, which are
well known
in the nonrelativistic scattering problem.
\par
Furthermore, the following reciprocity relations are proved to hold,
at least, up to the first order of CP
violation:
\eqn\bo{\eqalign{
T_{L \rightarrow L}^b + T_{R \rightarrow L}^b
&= T_{R \rightarrow L}^s + T_{R \rightarrow R}^s \bigl(= 1 - R_{R \rightarrow
L}^s \bigr), \cr
T_{L \rightarrow R}^b + T_{R \rightarrow R}^b
&= T_{L \rightarrow L}^s + T_{L \rightarrow R}^s \bigl(= 1 - R_{L \rightarrow
R}^s \bigr). \cr
}
}
The proof goes as follows. By inserting \bl~ into ${\tilde I_2}$
and using \bm, we have
\eqn\bp{\eqalign{
{\tilde I_2}={\beta^2 \over \alpha^2}
\bigl[&\gamma_{-}(-\alpha, \beta)\gamma_{+}(-\alpha, \beta)
I_2 -\gamma_{-}(-\alpha, \beta)\gamma_{+}(\alpha, \beta) I_1^*  \cr
-&\gamma_{-}(\alpha, \beta)\gamma_{+}(-\alpha, \beta) I_1
+\gamma_{-}(\alpha, \beta)\gamma_{+}(\alpha, \beta) I_2^*  \bigr].\cr
}
}
Remembering that $\gamma_{-}$'s are written in terms of $\gamma_{+}$'s as shown
in I,
%By making use of \bm~,
we arrive at
\eqn\bq{\eqalign{
\tilde \delta^{CP}=&-{\xi \over \epsilon}{\beta \over {2 \alpha}}
\bigl[{{\alpha - \xi} \over {\alpha}}
{{\gamma_{+}(-\alpha, \beta)} \over {\gamma_{+}(\alpha, \beta)}}
\bigl(\abs{\gamma_{+}(\alpha, \beta)}^2 - \abs{\gamma_{+}(-\alpha, \beta)}^2
\bigr)I_2 + c.c.
\bigr]  \cr
=&-{\xi \over \epsilon}{\beta \over {2\alpha}}
\bigl[{{\alpha - \xi} \over {\alpha }}
{{\gamma_{+}(-\alpha, \beta)} \over {\gamma_{+}(\alpha, \beta)}}{\alpha \over
\beta}I_2
+ c.c. \bigr] \cr
=&+{\xi \over {2 \epsilon}}{\beta \over \alpha}
\bigl[{{\gamma_{-}(-\alpha, \beta)} \over {\gamma_{+}(\alpha, \beta)}}I_2 +
c.c.
\bigr]=\delta^{CP}. \cr
}
}
Namely, the amount of CP violation is independent of the direction from which
the incident fermion comes. Eq.\bq~ combined with \bn~ leads to \bo~.
\par
We emphasize the fact that
various coefficients obtained here are expressed in terms of
only two quantities, $T^{(0)}$ and $\delta^{CP}$.
\newsec{Quantum-Number Flux through Bubble Wall}
\subsec{The flux into the symmetric phase}
Suppose that a left ( right )-handed fermion has a quantum number $Q_{L(R)}$,
which
is conserved in the symmetric phase.
Let us estimate the expectation value of the changes of the quantum number in
the
symmetric phase brought by a fermion incident from the symmetric phase.
It is given by
\eqn\nna{\eqalign{
\Delta Q^s &=(Q_R - Q_L )R_{L \rightarrow R}^s + (Q_L - Q_R )R_{R \rightarrow
L}^s +
(-Q_R + Q_L )\bar R_{L \rightarrow R}^s  \cr
&+(-Q_L +  Q_R )\bar R_{R\rightarrow L}^s
+(-Q_L)(T_{L\rightarrow L}^s + T_{L\rightarrow R}^s )+(-Q_R)( T_{R \rightarrow
L}^s +
T_{R \rightarrow R}^s )  \cr
&-( - Q_L) ( \bar T_{L \rightarrow L}^s + \bar T_{L \rightarrow R}^s ) -
(- Q_R)( \bar T_{R \rightarrow L}^s + \bar T_{R \rightarrow R}^s ) \cr
&=(Q_L - Q_R )(R_{R \rightarrow L}^s - \bar R_{R \rightarrow L}^s), \cr
}
}
where we have used the unitarity relations \au~.
On the other hand, that brought by a fermion from the broken phase is
\eqn\nnb{\eqalign{
\Delta Q^b =&Q_L(T_{L\rightarrow L}^b + T_{R \rightarrow L}^b ) +
Q_R ( T_{L \rightarrow R}^b + T_{R \rightarrow R}^b )   \cr
+& (- Q_L) ( \bar T_{L \rightarrow L}^b + \bar T_{R \rightarrow L}^b) +
(- Q_R) ( \bar T_{L \rightarrow R}^b + \bar T_{R \rightarrow R}^b )  \cr
=&-(Q_L - Q_R )(T_{L \rightarrow R}^b + T_{R \rightarrow R}^b - T_{L
\rightarrow L}^b
-T_{R \rightarrow L}^b).
}
}
{}From \nna~ and \nnb~,
the quantum-number flux into the symmetric phase just
in front of the bubble wall moving with velocity $u$ is given by
\eqn\ca{\eqalign{
F_Q&= {1 \over \gamma}\int_{m_0}^\infty dp_L \int^\infty_0 {dp_Tp_T
\over{4\pi^2}}
(Q_L-Q_R) \cr
     &\times \Bigl[(R^s_{R\rightarrow L}-{\bar R^s_{R\rightarrow
L}})f^s(p_L,p_T)  \cr
&-(T^b_{L\rightarrow R}+T^b_{R\rightarrow R}-T^b_{L\rightarrow
L}-T^b_{R\rightarrow L})f^b(-p_L,p_T)\Bigr], \cr
}
}
where $p_{L(T)}$ is the longitudinal ( transverse ) momentum of the fermion to
the
bubble wall,
$\gamma=\sqrt{1-u^2}$ and the fermion-flux density in the symmetric
(broken) phase $f^s(f^b)$ is given by
\ref\ckn{A. Cohen, D. Kaplan and A. Nelson, \NP{349} (1991) 727; \hfill\break
A. Nelson, D. Kaplan and A. Cohen, \NP{373} (1992) 453.}
\eqn\cb{\eqalign{
& f^s(p_L,p_T)=(p_L/E)\bigl(\exp {[\gamma (E-up_L)/T]}+1\bigr)^{-1}, \cr
& f^b(-p_L,p_T)=(p_L/E)\bigl(\exp {[\gamma (E+u{\sqrt
{p_L^2-m_0^2}})/T]}+1\bigr)^{-1}, \cr
}
}
with the chemical potential being omitted for simplicity.
Thanks to the reciprocity relations \bo~, \ca~ is reduced to a
simple expression:
\eqn\cc{
\eqalign{
F_Q= {1 \over \gamma}\int_{m_0}^\infty dp_L \int^\infty_0 {dp_Tp_T
\over{4\pi^2}}
(Q_L-Q_R)\Bigl[f^s(p_L,p_T)-f^b(-p_L,p_T)\Bigr]\Delta R.
}
}
Here $\Delta R$ is the difference between the chiral fermion and
its anti-fermion in the reflection coefficients incident from
the symmetric phase:
\eqn\cd{
\Delta R
\equiv R^s_{R\rightarrow L}-\bar R^s_{R\rightarrow L}
=-2T^{(0)}\delta^{CP},
}
where $T^{(0)}$ and $\delta^{CP}$ are given in \ar~ and \as~ respectively.
\par
We now recognize that $\Delta R$ is the dynamical quantity of primary
importance, since, if its absolute value is very small, any quantum-number
flow is almost forbidden. Note that the vector-like quantum numbers
such as baryon and lepton numbers do not flow through the
CP-violating bubble wall.
This statement would hold, at least, up to $O(g^2)$
\ref\fote{This statement may be exact because the reciprocity relations would
be expected to hold to all order of the perturbation theory with respect to
the CP violation.}.
The candidates for $Q_{L(R)}$ in the electroweak theory with usual
quantum-number
assignment for the matter fields are the hypercharge or
the third component of the weak isospin $T_3$.
In the charge-transport scenario in \ckn~, the hypercharge flux $F_Q$ carried
by the top
quark is subsequently converted into net baryon asymmetry by the sphaleron
transition.
\subsec{Numerical analyses of $\Delta R$: the kink-type bubble wall}
We have carried out numerical analyses of $\Delta R$ by taking
several forms of $g(x)$ \fotr~
assuming that the bubble-wall profile without CP violation is of the kink
type \ref\ayala{A. Ayala, J. Jalilian-Martin, L. McLerran and A. P. Vischer,
\PRD{49} (1994) 5559.}:
\eqn\da{
 f(x)=(1+\tanh x)/2.
}
The unperturbed solutions $\phi^{(\pm \a)}_\pm(x)$ in this background
are expressed in terms of the hypergeometric functions.
The transmission and reflection coefficients
without CP violation are
\eqn\db{
\eqalign{
 &T^{(0)}
={{\sin(\pi\a)\sin(\pi\b)}\over
  {\sin[{\pi\over2}(\a+\b+\xi)]\sin[{\pi\over2}(\a+\b-\xi)]}},\cr
 &R^{(0)}
={{\sin[{\pi\over2}(\a-\b+\xi)]\sin[{\pi\over2}(\a-\b-\xi)]}\over
  {\sin[{\pi\over2}(\a+\b+\xi)]\sin[{\pi\over2}(\a+\b-\xi)]}},\cr
}
}
respectively.  \par
The effects of CP violation can be evaluated, once the functional form
of $g(x)$ is given.
Note that the CP-angle can be removed by a constant bi-unitary
transformation for $g(x)=\Delta \theta f(x)$, where
the parameter $\Delta \theta$ characterizes the magnitude of
CP violation.
As a check, we have confirmed $\Delta R/\Delta\theta=0$ numerically in this
case.
We have studied the following examples of $g(x)$ with
various ranges of $dg(x)/dx$:
$$
\eqalign{
g(x)=&\Delta \theta f(x)^2,\quad  g(x)=\Delta \theta f(x)^3,\quad
g(x)=\Delta \theta f^{\prime}(x) ,  \cr
g^{\prime}(x) =&\Delta\theta\, {\rm sech} x,  \quad
g^{\prime}(x)= \Delta \theta\, {\rm sech}(2x), \cr
g^{\prime}(x)=&\Delta\theta\, {\rm sech}~(3x), \quad
g^{\prime}(x)=\Delta\theta\, {\rm sech} x \tanh x, \cr
g^{\prime}(x)=& \Delta \theta\, {\rm sech}(2x)\tanh x, \quad
g^{\prime}(x)=\Delta\theta\, {\rm sech}(3x)\tanh x. \cr
}
$$
Among these, $\Delta R/\Delta\theta$ for $g(x)=\Delta \theta f^{\prime}(x)$
is just $-2$ times that for
$g(x)=\Delta \theta f(x)^2$, since $f^{\prime}(x)=2f(x)-2f(x)^2$.
\par
The shape, magnitude and sign of $\Delta R/\Delta\theta$ are very sensitive to
the
functional form of $g(x)$. However the dependence of $\abs{\Delta
R/\Delta\theta}$ on
$m_0/a$
shows an interesting general trend. Namely, $\abs{\Delta R/\Delta\theta}$
decreases as $m_0/a$ increases when $m_0/a~\roughly{>}~1$, as far as our
numerical analyses are concerned. In other words,
$\abs{\Delta R/\Delta\theta}$ becomes smaller as the relevant fermion
is heavier for fixed $a$
when $m_0~\roughly{>}~a$.
In \Fig\fa{$\Delta R/\Delta \theta$ as a function of $E^*$ for various $a$, in
the case
where $g(x)=\Delta\theta f(x)^2$. The numerical values of $E^*$ and $a$
are given in the unit of $m_0$, the height of the bubble wall.}
and~\Fig\fb{$\Delta R/\Delta \theta$ as a function of $E^*$ for various $a$, in
the case
where $g(x)=\Delta \theta f(x)^3$. The numerical values of $E^*$ and $a$
are given in the unit of $m_0$, the height of the bubble wall.},
we have plotted the $E^*$ dependence of $\Delta R /\Delta \theta$ for several
inverse thickness of the bubble wall, $a$, in the cases
$g(x)=\Delta \theta f(x)^2$ and $g(x)=\Delta \theta f(x)^3$, respectively.
$\abs{\Delta R /\Delta \theta}$ in these figures is extremely small for
$m_0/a~\roughly{>}~2$.
We could say
that a heavier fermion with the larger Yukawa coupling does not always take
part
in baryogenesis game.
\newsec{Concluding Remarks}
We have proved that the reciprocity relations among
the transmission and reflection coefficients hold, at least, up to
the first order of CP violation.
All the coefficients are written in terms of only two
quantities, $T^{(0)}$ and $\delta^{CP}$, irrespective of whether an
incident fermion comes from the symmetric phase or the broken phase.
We have obtained the simple form of the quantum-number flux
through the CP-violating bubble wall with the help of the unitarity and
reciprocity.
If there is
difference in the quantum number between left- and right-handed fermions,
it can be left in the symmetric phase.
\par
The numerical analyses of $\Delta R / \Delta \theta$
indicate an interesting feature that the heavier fermion
plays a minor role to generate the baryon asymmetry when the Compton wave
length
is comparable with or less than the wall thickness. Specifically in the cases
$g(x)=\Delta \theta f(x)^2$ and $g(x)=\Delta \theta f(x)^3$,
$\abs{\Delta R}$ is negligibly small in the region $2/m_0 < 1/a$, which means
that twice the Compton wave length of the relevant fermion must be larger than
the wall
thickness in order for $\Delta R$ to have a sizable value. The
above feature seems to be reasonable, since the heavier fermion would
experience
thicker bubble wall due to its short Compton wave length. One cannot say that
the top quark
is the most relevant one contributing to the quantum-number flux $F_Q$.
{\it A larger Yukawa coupling does not necessary mean a larger effect of CP
violation
in electroweak baryogenesis}.
\par
The chemical-equilibrium relations in particle interactions in the
symmetric phase play an important role
to convert some quantum number, say the hypercharge, into the net baryon number
density by the sphaleron transition \ckn.
Yukawa interactions for light fermions like
$e$, $\mu$ and $u$ are out of chemical equilibrium in the symmetric phase,
because their
mean free times are comparable with the expansion time of the universe.
Thus the candidate fermions contributing to the baryon-number generation would
be
$\tau$ lepton and $d$, $c$, $s$, $t$ and $b$ quarks.
Once the wall thickness and the functional form of $g(x)$ are known, we can
select,
by using the formulae for $\Delta R$ and $F_Q$, what species of fermions take
part in
the baryogenesis game.
\par
Here we make a comment on
how to determine the CP-violating imaginary part $g(x)$ in two-Higgs-doublet
models.
One should note that $g(x)$ itself is not physical since it is a gauge-variant
quantity. The imaginary part $g(x)$ and the corresponding classical gauge-field
configurations would be determined by
solving the classical equations of motion of the gauge-Higgs
system in some gauge.
It is, however, difficult to solve the Dirac equation in the background of
classical
solutions of gauge and Higgs fields.   \par
Now, let us assume that the gauge fields are pure-gauge type:
\eqn\a{
ig{\tau^a \over 2}A_{\mu}^a=\del_{\mu}U_2(x)U_2^{-1}(x), \quad
ig^{\prime}{1 \over 2}B_{\mu}=\del_{\mu}U_1(x) U_1^{-1}(x),
}
where $U_1$ and $U_2$ are elements of $U(1)_Y$ and $SU(2)_L$ respectively.
We are interested in a bubble-wall solution which may be assumed to be
spherically symmetric. In this case we know an example that the pure-gauge
assumption
is reasonable.
As shown by Ratra and Yaffe in \ref\ry{B. Ratra and L. G. Yaffe, \PL{205}
(1988) 57.}
, the $SU(2)$ gauge-Higgs system can be reduced
to $1+1$ dimensional abelian gauge-Higgs system under the spherically symmetric
ansatz for the
gauge and Higgs fields. In that case the gauge field is written
in terms of pure-gauge.
Once we admit the assumption, the lagrangian of the gauge-Higgs system is
reduced to
\eqn\b{
\CL=\abs{\del_{\mu}\ph_1}^2 + \abs{\del_{\mu}\ph_2}^2 - V_{eff}(\ph_1, \ph_2
;T),
}
where $\ph_i (i=1, 2)$ are the Higgs fields in the pure-gauge
background. If we decompose $\ph_i$ as
${1 \over {\sqrt 2}} \rho_i (x){\rm e}^{i \theta_i (x)}$, the phases
$\theta_i(x)$
are unambiguously determined by the equations of motion, which follow from \b~.
Note that $V_{eff}$ depends only on $\theta_1 - \theta_2$ due to the original
gauge-invariance so that no $x$-dependent phase arises in the Higgs potential
in one-Higgs-doublet model. The equations of motion for $\theta_{i}$ are
\eqn\c{
\del_{\mu}(\rho_i^2 \del^{\mu}\theta_i) + {{\del V_{eff}} \over {\del
\theta_i}}=0,
\quad (i=1, 2).
}
{}From \c~, we have
\eqn\d{
\del_{\mu}(\rho_1^2 \del^{\mu}\theta_{1}+\rho_2^2 \del^{\mu}\theta_{2})=0
}
by using the fact that $V_{eff}$ depends only on $\theta_1 - \theta_2$.
On the other hand, the condition that the gauge fields stay in pure gauge
( the sourcelessness condition ) is given by
\eqn\e{
\rho_1^2(x)\del_{\mu}\theta_1(x) + \rho_2^2(x)\del_{\mu}\theta_2(x) =0,
}
which follows from the equation of motion for the gauge fields.
It is important that we should generally
impose \e\  rather than \d\  when we solve classical equations of motion
under the pure-gauge assumption for gauge fields.
{}From the phases $\theta_i$ determined in this way, one can apply the
prescription
to solve the Dirac equation with CP-violating bubble wall background developed
in I and
this paper.
\par
Although many complicated factors
would affect baryogenesis, a sufficient amount of baryon-number
asymmetry could not be produced unless $\abs{\Delta R/\Delta \theta}$ due
to CP violation is large enough.
We expect that our DWBA prescription and the results would serve to
build models to generate baryon asymmetry of the universe.
%%%%%%%%%%%%%%%%%%%%%% Acknowledgements %%%%%%%%%%%%%%%%%%%%%%%%%%%%%%%%%%%%%%
%
%\bigbreak\bigskip
%\centerline{{\bf Acknowledgments}}\nobreak
%The authors would like to express their cordial gratitude to colleagues at
%%Saga, Kyushu
%and Kinki Universities for discussion and encouragement.
%%%%%%%%%%%%%%%%%%%%% References %%%%%%%%%%%%%%%%%%%%%%%%%%%%%%%%%%%%%%%%%%%%

\bigskip\bigskip\footatend\vfill\immediate\closeout%
\rfile\writestoppt\baselineskip=14pt\centerline{{\bf References}}%
\nobreak\bigskip{\frenchspacing%
\parindent=20pt\escapechar=` \input refs.tmp\vfill}\nonfrenchspacing
\bigskip\bigskip\vfill\immediate\closeout%
\ffile{\parindent40pt\baselineskip14pt\centerline%
{{\bf Figure Captions}}\nobreak\medskip
\escapechar=` \input figs.tmp\vfill\eject}

\bye